# Portable Silent Room: Exploring VR Design for Anxiety and Emotion Regulation for Neurodivergent Women and Non-Binary Individuals

Kinga Skierś ⓘ, Yun Suen Pai ⓘ, Marina Nakagawa ⓘ, Kouta Minamizawa ⓘ, Giulia Barbareschi ⓘ

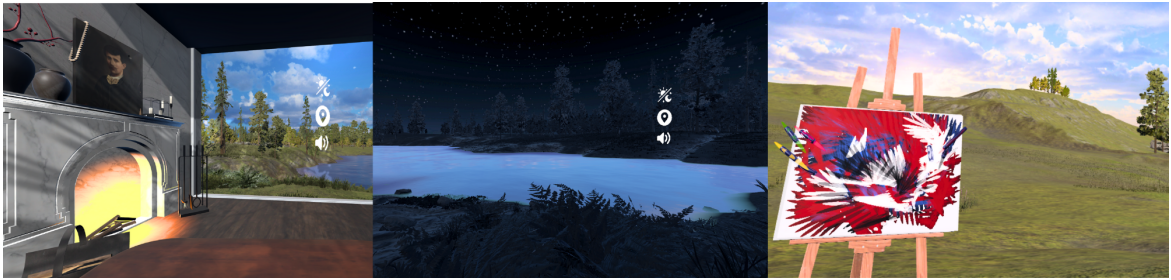

Fig. 1: Left: Indoor 'Living Room' environment Center: River Side during Night Right: Meadow with painting activities.

**Abstract**—Neurodivergent individuals, particularly those with Autism and Attention Deficit Hyperactivity Disorder (ADHD), frequently experience anxiety, panic attacks, meltdowns, and emotional dysregulation due to societal pressures and inadequate accommodations. These challenges are especially pronounced for neurodivergent women and non-binary individuals navigating intersecting barriers of neurological differences and gender expectations. This research investigates virtual reality (VR) as a portable safe space for emotional regulation, addressing challenges of sensory overload and motion sickness while enhancing relaxation capabilities. Our mixed-methods approach included an online survey (N=223) and an ideation workshop (N=32), which provided key design elements for creating effective calming VR environments. Based on these findings, we developed and iteratively tested VR prototypes with neurodivergent women and non-binary participants (N=12), leading to a final version offering enhanced adaptability to individual sensory needs. This final prototype underwent a comprehensive evaluation with 25 neurodivergent participants to assess its effectiveness as a regulatory tool. This research contributes to the development of inclusive, adaptive VR environments that function as personalized "portable silent rooms" offering neurodivergent individuals on-demand access to sensory regulation regardless of physical location.

✦

## 1 INTRODUCTION

*Neurodiversity* refers to the inherent variation in neurological functioning across individuals within a population [64]. This term explains that neurological variety in cognitive and sensory functioning, as seen in conditions like Attention-Deficit/Hyperactivity Disorder (ADHD), Autism, Dyslexia, and other cognitive variations, is a natural and valuable aspect of human diversity [12]. Neurodiversity challenges the traditional view of neurological differences as deficits that must be normalized to fit the 'neurotypical' majority [64], instead advocating for the recognition of these differences as comparable to other forms of diversity such as ethnicity, gender, and sexual orientation [13]. In this paper, we use the term neurodivergence to encompass conditions such as Autism and ADHD, highlighting the broad spectrum of neurocognitive patterns that exist across individuals. The concept of neurotypicality itself remains contested, with unclear boundaries between what is considered typical or atypical functioning, suggesting that human diversity in mind and behavior is vast and shaped by cultural and historical contexts [64]. Despite theoretical frameworks emphasizing inclusion, neurodivergent individuals frequently encounter environmental and societal barriers that significantly impact their mental well-being [49]. Research shows how adults with ADHD and autism experience elevated risks of anxiety, emotional dysregulation, sensory overload, and difficulties in self-regulation, often resulting in higher rates of depression and suicidal ideation [59]. These challenges are particularly pronounced for neurodivergent women and non-binary individuals, who must navigate the intersection of neurological differences, gender roles, and social expectations [36]. In recent years, research examining the role of technology to support mental health for neurodivergent individuals has gained increasing attention [77]. Prior studies have explored smartwatch applications to support concentration and reduce stress [14], online gaming communities as spaces for belonging [62], and the critical analysis of technological trends led by neurodivergent individuals themselves [69]. However, a review by Spiel et al. [69] identified a critical gap: most technological interventions focus on children or parenting contexts, with only a fraction designed for adults. In addition, many existing approaches aim to correct neurodivergent behaviors rather than provide supportive spaces aligned with neurodivergent needs [52].

This study investigates how Virtual Reality (VR) can be leveraged to create a Portable Silent Room, an adaptive safe space that helps neurodivergent individuals regulate emotions, recover from sensory

- *Kinga Skierś is with Keio University, Graduate School of Media Design. E-mail: kinga.skiers@gmail.com*
- *Yun Suen Pai is with Inclusive Reality Lab, School of Computer Science, University of Auckland. E-mail: yun.suen.pai@auckland.ac.nz*
- *Marina Nakagawa is with Keio University, Graduate School of Media Design. E-mail: maripasta.marina@gmail.com*
- *Kouta Minamizawa is with Keio University, Graduate School of Media Design. E-mail: kouta@kmd.keio.ac.jp*
- *Giulia Barbareschi is with Research Center Trustworthy Data Science and Security, University of Duisburg-Essen, Germany E-mail: giulia.barbareschi@uni-due.de*

overload, and mitigate anxiety. We explore key design elements necessary to ensure accessibility, comfort, and effectiveness while minimizing potential triggers such as motion sickness or overwhelming stimuli. Through thematic analysis of an online survey (N=223) and an ideation workshop (N=32) with neurodivergent adults, we identified core system components that contribute to a calming VR environment. Based on these insights, we developed and tested two VR prototypes in an initial study with 12 neurodivergent women and non-binary individuals, then iteratively refined the system before conducting a final evaluation with 25 neurodivergent women and non-binary participants to assess its effectiveness. The following research questions guide this work:

- **RQ1:** What design principles are essential for creating a VR-based Portable Silent Room tailored to the needs of neurodivergent women and non-binary individuals?

- **RQ2:** What environmental and interaction elements contribute to emotional regulation and relaxation in a VR Portable Silent Room, and which features may be overwhelming or inadequate?

- **RQ3:** How can VR systems more broadly be adapted to accommodate individual sensory preferences and enhance accessibility for neurodivergent users?

By addressing these questions, this research contributes to developing more inclusive and adaptive VR environments, offering insights into how technology can be designed to effectively support neurodivergent individuals in managing their emotional well-being. Our findings challenge existing assumptions about VR accessibility for neurodivergent users and provide concrete design recommendations for creating immersive experiences that address the unique wellbeing needs of this underserved population.

## 2 Related Works

### 2.1 Neurodiversity: ADHD and Autism

Attention-Deficit/Hyperactivity Disorder (ADHD) and Autism are complex neurodevelopmental conditions that significantly impact individuals' daily functioning and well-being. ADHD manifests through persistent patterns of inattention, hyperactivity, and impulsivity that deviate from typical developmental trajectories [44]. Autism encompasses a constellation of traits, including challenges in social communication and interaction, repetitive behavioral patterns, and intense, focused interests [24]. These neurological variations frequently affect adaptive functioning across multiple domains including academic achievement, occupational performance, and social relationships. Adults with ADHD often encounter substantial challenges in workplace environments and daily life management compared to neurotypical peers. These difficulties typically include sustained attention, time perception, task organization, and adherence to sequential instructions [4]. Autistic individuals frequently experience challenges interpreting non-verbal communication cues, developing and maintaining relationships, and navigating complex social dynamics [51] often alongside repetitive physical behaviors (stimming) or intense focus on specific subject areas. Significantly, both ADHD and Autism commonly co-occur with additional conditions such as anxiety disorders, depression, sensory processing differences, and misophonia, creating multifaceted experiences that extend beyond primary diagnostic criteria [22, 23]. The increasing recognition that these conditions exist along continuous spectra rather than as discrete categories have led to a more nuanced understanding of neurodivergent experiences, including acknowledgment of the substantial variation in both challenges and strengths across individuals.

### 2.2 Gender Equality in Diagnosis and Experience

Historically, both ADHD and Autism have been characterized as conditions predominantly affecting males, yet contemporary research reveals that neurodivergent women are significantly more prevalent than previously recognized [31, 36]. The diagnostic criteria and clinical interpretations of neurodivergent behaviors often result in systemic under-diagnosis or misdiagnosis in women and gender-diverse individuals. Traits such as verbal expressiveness or emotional responsiveness in women are frequently misattribute to personality characteristics rather than recognized as ADHD or Autism manifestations. Similarly, symptoms including executive dysfunction, organizational challenges, or anxiety present with different phenotypic expressions across genders that current diagnostic frameworks fail to adequately capture [60]. Longitudinal research demonstrates that girls with ADHD face disproportionate academic challenges, social difficulties, and elevated risks of comorbid conditions, including eating disorders, depression, and self-injurious behaviors compared to neurotypical peers [33, 57]. Autism, similarly, was historically conceptualized as a male-dominant condition, though increasing recognition of diverse presentations is gradually reshaping clinical understanding [20]. Despite this evolution, many autistic women continue to experience considerable barriers to accurate diagnosis and appropriate intervention. These diagnostic delays significantly impact mental health outcomes, educational achievement, employment opportunities, and interpersonal relationships [1]. Qualitative research by Kanfiszer et al. [32] examining the lived experiences of seven women diagnosed with autism in adulthood revealed consistent patterns of social interaction difficulties, negative peer experiences, and the development of complex camouflaging behaviors to meet normative social expectations. Women who remain undiagnosed frequently encounter persistent daily challenges attributable to lacking the conceptual framework and accommodations that appropriate diagnosis could provide [80]. Post-diagnosis experiences often include significant improvements in self-understanding and enhanced capacity to articulate specific needs and access appropriate supports [9]. Beyond binary gender classifications, non-binary and gender-diverse individuals face compounded challenges in receiving accurate diagnoses and appropriate support. While research specifically addressing non-binary neurodivergent experiences remains limited, emerging evidence suggests a significantly higher prevalence of neurodivergence among gender-diverse populations compared to cisgender cohorts [19, 75].

The intersection of gender diversity and neurodivergence creates unique psychological burdens, as societal pressures to conform to gender norms compound the challenges of navigating neurodivergent traits. Non-binary individuals frequently experience heightened psychological distress from simultaneously masking both gender identity and neurodivergent characteristics when navigating normative social contexts, contributing to elevated rates of anxiety, depression, and related mental health conditions [78]. Current diagnostic frameworks and research methodologies remain predominantly derived from studies involving cisgender male participants, creating substantial gaps in understanding the diverse presentations, unique needs, and adaptive strategies of neurodivergent individuals across the gender spectrum, particularly those with non-binary identities [25, 72, 73]. Neurodivergent women and non-binary individuals frequently encounter barriers when attempting to adapt to societal expectations, particularly within educational institutions, workplace environments, and social contexts that are predominantly designed for neurotypical cognitive styles, and according to traditional gender norms [12, 19]. These barriers often contribute to heightened stress levels, emotional dysregulation, and sensory overwhelm.

### 2.3 Emotional Regulation and Neurodiversity

Emotional self-regulation encompasses an individual's capacity to consciously and adaptively modulate affective responses to achieve goal-directed behaviors within socially appropriate parameters. This complex cognitive-emotional process fundamentally underpins psychological well-being and effective interpersonal functioning. Emotional dysregulation manifests across a multidimensional spectrum with individuals experiencing variable difficulties in emotional modulation [47, 66, 68].
Anxiety, beyond its psychological dimensions, frequently presents with physiological manifestations including cardiovascular arousal and hyperhidrosis [70]. Neurodivergent individuals demonstrate particular vulnerability to anxiety and mood dysregulation [10, 16], with emotional reactivity often triggered by interpersonal stressors such as evaluative feedback in professional or educational contexts, or relationship challenges [37, 76]. Neurophysiological research demonstrates that sensory processing pathways in neurodivergent individuals differ significantly from neurotypical populations, creating substantial challenges in navigating everyday environments [2, 53].

Sensory processing differences present along a bidirectional spectrum: some individuals experience hypersensitivity to environmental stimuli, resulting in attention regulation difficulties and sensory-seeking behaviors [81], while others encounter hypersensitivity leading to sensory overload when exposed to excessive stimulation. These overload states trigger distress episodes characterized by both physiological and psychological manifestations of acute stress [43]. Environmental control emerges as a critical component of emotional regulation for neurodivergent individuals. The capacity to modulate sensory input either through amplification or attenuation of stimulus intensity and access to spaces permitting sensory customization is essential for preventing overwhelming experiences, facilitating emotional equilibrium, and supporting longitudinal well-being [46, 53]. While individual adaptive strategies such as acoustic filtering devices provide partial mitigation, comprehensive environmental modifications (e.g., illumination adjustments or acoustic treatments) present greater implementation challenges, particularly in shared environments. Furthermore, physical and social barriers frequently prevent neurodivergent individuals, particularly women and non-binary people, from accessing dedicated safety zones where they can effectively manage sensory experiences. These accessibility limitations negatively impact both immediate emotional regulation capacity and overall quality of life [53, 81]. While VR shows promise for emotional regulation, most systems are based on studies with cisgender male participants, limiting their relevance for diverse neurodivergent users. Recent critiques in HCI highlight a lack of gender inclusion in design and evaluation. For example, Riches et al. [61] report that only one VR relaxation study included non-binary participants, with too small a sample to analyze, underscoring persistent underrepresentation. For autistic women and non-binary individuals, emotional challenges are often shaped by masking, underdiagnosis, and social pressure [11, 61]. Biological factors also matter: PMDD-related shifts in blood flow and sensory sensitivity can disrupt emotional stability, introducing regulation needs rarely addressed in existing systems [27].

### 2.4 Virtual Reality and 'Safe Spaces'

Virtual Reality has emerged as a transformative modality in personal wellbeing applications, offering unprecedented approaches to develop personalized interventions for emotional and psychological support [35, 67]. By facilitating user immersion in controlled, responsive environments, VR enables precisely calibrated interventions adaptable to individual sensory and emotional requirements [29]. A particularly promising application domain is the implementation of virtual safe spaces personalized, immersive environments specifically engineered for emotional regulation and sensory modulation. These environments offer particular benefits for neurodivergent individuals who might struggle to access supportive physical environments [7, 30]. VR safe spaces can be parametrically adjusted to mitigate sensory overload, integrate structured coping mechanisms, and enhance perceived environmental control.

Despite demonstrating significant potential as a platform for sensory-informed interventions, current VR research and design methodologies remain constrained in their inclusivity, particularly regarding neurodivergent women and gender-diverse populations. A substantial gender disparity characterizes VR research participation, with male subjects outnumbering female participants at a ratio of 3:1 [48]. Additionally, numerous VR applications fail to adequately address the elevated susceptibility to cybersickness and sensory discomfort frequently reported by neurodivergent users [5, 42]. While specific research initiatives have targeted cybersickness reduction [34, 54] and sensory overload mitigation in virtual environments [39], these approaches have rarely been integrated into comprehensive design frameworks. Recent clinical investigations have demonstrated the therapeutic efficacy of VR calm rooms in psychiatric contexts, documenting improvements in subjective well-being and anxiety reduction through immersive, non-pharmacological interventions [28]. Similarly, VR sensory rooms have shown effectiveness in reducing anxiety and enhancing sensory processing capabilities among adults with various disabilities [50]. Moreover, comparative studies indicate that VR-mediated mindfulness practices demonstrate superior efficacy for stress reduction compared to traditional methodologies, highlighting the potential of personalized VR interventions [21]. However, existing VR sensory-modulated environments for emotional regulation have not been systematically designed with specific consideration for neurodivergent women and non-binary individuals, creating a significant gap in both theoretical frameworks and practical applications. By prioritizing the development of VR safe spaces explicitly tailored to accommodate the unique needs of neurodivergent women and gender-diverse individuals. Our research aims to address this critical disparity and advance more effective, inclusive, and accessible virtual environments for emotional regulation and psychological well-being across diverse neurodivergent populations. We do this by first capturing essential design requirements from direct engagement with users and iteratively develop a Portable Silent Room, a customizable VR space that allows participants to manipulate the environment around them to manage sensory stimulation as well as engage in relaxing activities to facilitate emotional self-regulation.

## 3 METHODOLOGY

Our research lasted fourteen months and was driven by the first author's exploration of their neurodivergent identity and the challenges faced in diagnosis. This experiential foundation, complemented by extensive engagement with neurodivergent communities, revealed significant mental health challenges impacting daily functioning. Our choice was to prioritize the representation of neurodivergent women and non-binary individuals which are less frequently targeted by VR and other technology-centered design projects. Our goal was to support agency and participation among neurodivergent individuals who are more often overlooked. One key decision in this process was to extended recruitment to those who had completer a process of self-diagnosis using validate tools. Such approach is consistent with existing scholarship acknowledging structural barriers to diagnosis, particularly for women and gender-diverse individuals, such as high costs, diagnostic delays, lack of language access, and widespread gender bias in assessment [3, 56]. Throughout the research, we adopted a structured co-design approach to address the spectrum of needs across system features, including environmental aesthetics, auditory elements, visual parameters, spatial configuration, and interactive environments. Due to the complexity of the task and the lack of resources that documented the specific requirements of neurodivergent women and non-binary persons with regards to the use of VR as well as the practices they engage with in order to manage sensorial needs and promote relaxation, we utilized a multi-step approach including:

1. Formative study: consisting of an initial online survey to explore sensory preferences, followed by a series of one-time individual or small-group workshops (based on participant preference) conducted over eight weeks, aimed at defining core requirements for the VR Portable Silent Room.

2. Initial Prototype Design and Testing: based on the insights from the Formative study, we developed an initial prototype composed of two different VR environments with different features which were evaluated by 12 neurodivergent women and non-binary participants.

3. Final Prototype Design and Testing: utilizing the feedback provided by participants, we created the final version of the Portable Silent Room which includes six separate scenes, options for engaging in creative and breathing activities, and a control panel to regulate light, audio and environment.

All studies received ethics approval from the University's ethics comitee. A diagram showing the steps of our methodological approach is shown in Figure 2.

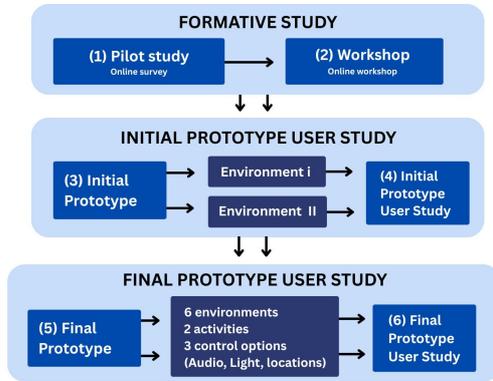

Fig. 2: Diagram of methodology

## 4 FORMATIVE STUDY

### 4.1 Methods

As an initial step, we conducted an anonymous online survey targeting adults with ADHD and Autism to understand their relaxation practices and sensory preferences. This was followed by a series of eight workshops with women and non-binary neurodivergent individuals to define design requirements for a VR system that could be used as a portable safe space.

#### 4.1.1 Participants

The online survey was distributed in several international social media neurodivergent groups. Workshop participants were recruited through the same social media neurodivergent groups as well as through neurodivergent-related networking channels. A total of 223 individuals provided completed responses to the survey. Participant demographics included age distribution (18-24: 32.3%, 25-34: 43.8%, 35-44: 16%, 45-54: 7%) and gender (Female: 72%, Male: 12.6%, Non-binary: 15.4%). Following the analysis of the survey results, we conducted workshops over a period of eight weeks with a total of 22 women and 10 non-binary individuals, all of whom were either self- or officially diagnosed with autism and/or ADHD. Each participant took part in a single workshop session, which ranged from one to three hours in duration. Based on individual preferences and anonymity needs, participants could choose to join either a group session or a one-on-one session. The workshops were held online, as participants were based in various countries.

#### 4.1.2 Procedure

The online survey included questions about respondents' ongoing mental health challenges emotional self-regulation strategies, sensory processing self-management, and stress relief activities.

Building on initial insights, we developed a framework for the workshop in order to explore in more depth sensory triggers, emotional regulation strategies, and user preferences for a VR-based system which could support relaxation and emotional regulation. Sessions were facilitated using MiroBoard[1], where participants engaged in group or individual activities. Each participant had their own board and completed activities focused on: Identifying personal triggers, sharing music and sound preferences, designing and co-creating definition of 'safe space', and discussing desired features and functions for a VR-based relaxation system; including aspects such as color schemes, environment design, lightning preferences, and presence or absence of auditory elements. Discussions were recorded with consent from participants.

[1] https://miro.com/

#### 4.1.3 Data Analysis

Survey responses were analyzed to identify patterns in participants' experiences, challenges, and coping strategies. Workshop data was collected through MiroBoard platform and recorded discussions. Recordings from the workshops were transcribed by the first author. An initial round of coding was conducted manually by the first author which assigned labels to relevant quotes recordings and notes from participants' boards. Affinity diagrams and progressive aggregation were then used to and identify common themes [41], with the authors engaging in repeated discussions during online meetings to consolidate a comprehensive interpretation of the dataset.

### 4.2 Results

Analysis of survey responses (N=223) revealed significant mental health challenges among participants. Over 65% reported experiencing anxiety, or depression within the two weeks prior to the survey, while only 13% reported no such experiences. More than 40% of respondents indicated at least one co-occurring mental health condition affecting their daily functioning, with Depression (N=60) being most prevalent, followed by Misophonia (N=27), Sensory Processing Disorder (N=23), and Auditory Processing Disorder (N=20). Regarding stress management strategies, participants reported both passive and active approaches. Passive engagements included listening to music (N=151) and watching movies or television (N=103), while active involvements; creative activities such as gaming or drawing (N=102) and spending time in nature (N=82). These findings highlighted the need for a balanced approach accommodating strategies for both engagement seeking and stimulation avoidance, in the design of tools for relaxation. Similarly to what was documented in the survey, workshop participants consistently described struggles with sensory processing, executive functioning, and emotional regulation that significantly impacted their daily lives, relationships, and career trajectories. Through affinity diagramming, we identified three primary categories of challenges:

**Environmental Stressors:** Participants frequently cited workplace and educational contexts as particularly challenging, with common triggers including strict deadlines, information overload, complex social expectations, work pressure, and lack of privacy in open office environments.

**Sensory Triggers:** Specific sensory stimuli elicited strong negative responses, with auditory triggers being particularly prominent, including loud or repetitive sounds (chewing, overlapping conversations), sudden lighting changes, and certain tactile sensations from clothing or furniture materials.

**Emotional Reactivity:** Many participants described intense emotional responses, including anxiety, overwhelm, and depressive episodes, leading to temporary cognitive paralysis or social withdrawal. These reactions often require extended recovery periods and disrupt daily functioning.

When discussing personal *Safe Space* characteristics, many participants emphasized nature elements such as forests, oceans, and mountains, seeking escape from human overstimulation into natural environments. Safe spaces weren't limited to outdoor environments; participants mentioned various indoor spaces ranging from bedrooms to kitchens and bathrooms. They emphasized the importance of feeling 'at home', with multiple individuals noting they had suffered from lack of access to their own personal space in the real world. Notably, a safe space could be characterized by absence of various elements as much as their presence. A significant portion of users express interest towards 'empty space,' described as open, dark, and quiet, with some specifically mentioning the presence or distant features such as stars or galaxies, and others had an interest in the use of biofeedback visualization, particularly hearth beat, to promote a better sense of connection with the body from which they often felt disconnected during stressful episodes.

Participants consistently reported that existing support mechanisms, including therapy and medication, provided insufficient tools for managing these challenges in real-time environments. The expressed need was for accessible tools that could provide immediate sensory and emotional regulation support without requiring extensive preparation or specific environments and which one could access independently of their physical location.

## 5 INITIAL PROTOTYPE USER STUDY

### 5.1 Design

Our VR system was developed based on insights gained from a the Formative Study. We utilized the Meta Quest 2[2] VR headset with dual controllers for navigation, BOSE QuietComfort 45 headphones for noise cancellation, Emotibit[3] sensor for physiological data capture, and an Alienware m18 R2 for computational processing. Both prototypes were implemented in Unity3D[4](ver. 2022.3.20f1, with file size 3.50 GB) with custom C# scripting. Thematic analysis of participant data revealed two distinct experiential preferences: (1) grounded, nature-inspired settings and (2) abstract, immersive spatial environments. Audio preference analysis similarly identified two primary categories: nature soundscapes (e.g., hydroacoustic, avian, used in *Environment I*) and frequency-modulated abstract audio (e.g., brown noise variants, used in *Environment II*). These findings guided the development of two distinct environments:

**Environment I: Immersive Nature and House.** This implementation featured a procedurally generated natural landscape with topographical variation (mountains, forests) and an architecturally optimized structure for controlled exploration. Users navigated via spherical teleportation nodes positioned at spaced intervals throughout the environment. A guided respiratory exercise module was integrated in a dedicated virtual space. The architectural interior was designed according to participant-provided reference imagery, emphasizing natural illumination, biophilic elements, and visual coherence.

**Environment II: Galaxy and Biofeedback Experience.** This prototype featured a static environment with a celestial skybox and low-illumination mountainous terrain. Biofeedback integration was achieved through a reactive spherical element that modulated movement parameters based on heart rate variability (HRV). The Emotibit sensor captured HRV via photoplethysmography, processed through TouchDesigner[5] middleware, and transmitted to Unity3D via UDP protocol. The biofeedback implementation was guided by previous research demonstrating physiological state visualization as an effective mechanism for interoceptive awareness and emotional regulation [67, 79].

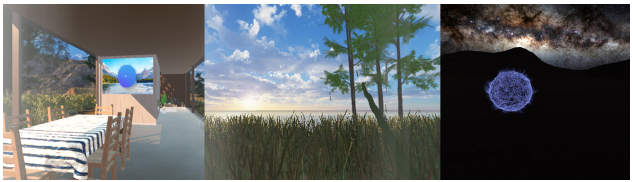

Fig. 3: Left: Prototype I: Immersive Nature and House, Right: Prototype II: Galaxy and Biofeedback Experience

### 5.2 Methods

#### 5.2.1 Participants

The evaluation involved 9 female and 3 non-binary participants (ages 18-44). Eight participants had clinical ADHD diagnoses, three were diagnosed with Autism, and one self-reported ADHD symptoms alongside comorbid Depression and Anxiety Disorder. Half of the participants had prior experience using VR, while the remaining six had no previous exposure. All participants were recruited through neurodivergent advocacy groups, online communities, and professional networks.

[2]https://www.meta.com/quest/products/quest-2/
[3]https://www.emotibit.com/
[4]https://unity.com/
[5]https://derivative.ca/

| ID | Gender | Age | VR Exp. | Diagnosis | Dx Method |
|---|---|---|---|---|---|
| P1 | NB | 26–34 | Yes | ADHD | Diagnosed |
| P2 | F | 26–34 | Yes | ADHD | Diagnosed |
| P3 | F | 26–34 | Yes | Autism | Diagnosed |
| P4 | NB | 26–34 | No | ADHD | Diagnosed |
| P5 | NB | 18–25 | Yes | ADHD | Diagnosed |
| P6 | F | 18–25 | Yes | ADHD | Diagnosed |
| P7 | F | 26–35 | Yes | ADHD | Diagnosed |
| P8 | F | 18–25 | No | ADHD | Self-reported |
| P9 | F | 35–39 | No | Autism | Diagnosed |
| P10 | F | 40–44 | No | ADHD | Diagnosed |
| P11 | F | 35–39 | No | ADHD | Diagnosed |
| P12 | F | 35–39 | No | Autism | Diagnosed |

Table 1: Participant demographics, VR experience, and diagnosis method. F = Female, NB = Nonbinary. Dx = Diagnosis. AuADHD=Autism & ADHD

#### 5.2.2 Procedure

The study was conducted in a university setting. After signing consent forms acknowledging potential motion sickness risks and withdrawal rights, participants completed a pre-survey questionnaire. Participants created a personalized relaxation space using provided cushions and weighted blankets. Once comfortable, they wore a VR headset and noise-cancelling headphones for the *Environment I: Immersive Nature and House* (5 minutes), followed by a brief post-experience survey and interview. Next, the EmotiBit sensor was attached to their left index finger for the *Environment II: Galaxy and Biofeedback* (5 minutes). They then completed a final survey and semi-structured interview on system usability, immersion, and relaxation effects. All participants experienced the environments in the same order, beginning with the lighter nature scene before transitioning to the darker Galaxy experience based on pre-study preferences for dimmer environments. The exploratory design allowed participants to freely engage without assigned tasks.

#### 5.2.3 Data Analysis

The data analysis focused primarily on the custom survey responses regarding personal experience and satisfaction ratings. We analyzed the interview data about neurodiversity and experience, identifying what participants liked and what challenges they encountered. Interview recordings were transcribed by the first author. Transcripts and survey responses were then manually coded and discussed. Analysis of feedback was initially conducted separately for each environment to assess their impact on participants. Following this, inductive thematic analysis [8] was leveraged to conceptualize themes to extract broader design consideration through discussions among the authors until consensus was reached.

### 5.3 Results

The analysis from the data related to the *Environment I* revealed that participants responded positively to the environment's exploratory freedom and personalization options. Natural elements, particularly water features and avian audio were consistently identified as effective relaxation catalysts. However, several usability challenges emerged, including unintentional teleportation, collisions with virtual boundaries, which significantly disrupted immersion continuity and caused mild motion sickness in some participants. Multiple participants expressed fatigue with the step-by-step movement mechanics. The data analysis related to to *Environment II* analysis showed more polarized responses. A subset of participants reported heightened engagement and immersion, while others found the low-illumination cosmic environment disorienting and anxiety-inducing. The biofeedback component introduced physical interference through the finger-mounted sensor that counteracted relaxation for certain participants.

Participant feedback revealed several critical design considerations for neurodivergent-focused relaxation environments:
**Environmental Adaptability:** Users expressed strong preferences for environments responsive to their current cognitive and sensory states, particularly regarding illumination levels. While some found Prototype II's darkness discomforting, others reported Prototype I's brightness as overstimulating, indicating the necessity for adjustable lighting parameters.
**Audio Customization:** Default soundscapes received mixed responses, highlighting the importance of providing user control over both audio selection and volume calibration to accommodate individual auditory sensitivities.
**Navigation Refinement:** The movement-dependent navigation system was frequently identified as attention-diverting. Participants predominantly preferred stationary relaxation experiences with minimal navigation requirements, citing potential disorientation and motion discomfort as concerns.
**Engagement Balance:** Rather than purely passive experiences, participants expressed the desire for optional lightweight interactive elements that provide agency without demanding cognitive effort, praising for example the presence of optional breathing exercises, which created balanced relaxing engagement.

These insights guided the development of our final prototype, which incorporated refined interaction mechanics, expanded environmental controls, and responsive design elements. Compared to past VR tools that are driven by male-dominant preference, these insights are unique to neurodivergent women and gender-diverse individuals in terms of their sensory sensitivity level, cognitive load and interface accessibility.

# 6 FINAL PROTOTYPE USER STUDY
## 6.1 Design

Building upon insights from the previous study, our final Portable Silent Room prototype integrates multi-environment immersion within a cohesive spatial framework, featuring a modern residential structure in a procedurally generated natural landscape. Using the same Unity setup as previous prototypes, this implementation emphasizes accessibility, interaction fidelity, and customizable sensory engagement, while maintaining cross-platform compatibility and real-time environment adaptation.

**Environmental and Spatial Design:** The system renders a comprehensive open-terrain environment (4000m²) with dynamic level-of-detail scaling to maintain optimal performance. At its core stands a transparent facade structure with generated glass walls utilizing real-time reflection and refraction shaders to create unobstructed sightliness to the surrounding ecosystem. The architectural implementation comprises three distinct functional zones:
(1)Living Room: Features a central fireplace with simple fire effects that create a cozy, warm atmosphere. The gentle light and subtle animations provide a visual focal point for relaxation.
(2)Bedroom: Designed as a quiet retreat space with soft lighting and minimal distractions. The enclosed design creates a sense of safety and security for users seeking a more sheltered environment.
(3)Kitchen: Contains a television display showing guided breathing exercise animations. Users can follow along with these visual cues to practice structured breathing techniques that support relaxation.
The exterior environment features three distinct outdoor areas, each designed to offer different relaxation experiences:
(1) Forest: A nature environment featuring dense vegetation that creates spatial enclosure, designed to facilitate forest bathing therapeutic practices. Dynamic foliage animation enhances environmental fidelity and provides ambient visual feedback.
(2) Meadow: An open, expansive area with clear sightliness and minimal obstacles, creating a sense of freedom and openness. The lighting system enhances depth perception through subtle atmospheric effects. This environment features a Creating Expression Module which enables participants to paint on a blank canvas to support emotional regulation through more active engagement.

(3)Riverside: Located between the lake and surrounding vegetation, this transitional space features gentle water animations with realistic reflections and ripples, balancing the calming effects of water with the security of nearby landscape features. These environments were designed using standard 3D modeling techniques and optimized terrain systems to ensure smooth performance while maintaining visual quality appropriate for relaxation and emotional regulation purposes. Each area was specifically crafted to address different sensory preferences identified in our user studies.

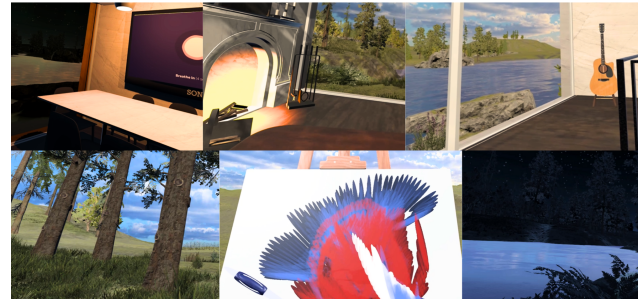

Fig. 4: Final prototype, From Up left: Kitchen, Living Room, Bedroom. From Down Left: Forest, Meadow, Riverside

**User Interface and Interaction Framework:** To address previously identified usability constraints, we developed a spatially anchored interface with multimodal interaction capabilities. This allows users to choose between day and night settings, as well as different virtual locations, providing more control over their environment and helping to reduce anxiety and motion sickness previously caused by limited options in the prior system.
**Location Transition System:** A six-node teleportation network, allowing users to freely select and transition between any of the six available locations through the user interface.
**Diurnal/Nocturnal Transition:** Implementation of a real-time global illumination controller that dynamically adjusts ambient light sources, shadow parameters, and color temperature. The system maintains minimum visibility thresholds during nocturnal settings with warm-spectrum illumination sources strategically placed for optimal spatial awareness without disrupting relaxation.
**Auditory Stimulus Management:** Based on the results gathered from previous user studies and workshops [55, 63], we implemented six distinct audio scientifically associated with neurodivergent relaxation responses: Black and Pink noise, Lo-Fi composition, Night-mode ambient compositions, Operatic selections, Classical piano repertoire, and Nature-based recordings with spatial audio implementation. All audio-profiles are volume adjustable.
**Breathing Practice Interface:** In the Kitchen scene, the system implements two visual respiratory guidance algorithms displayed on a virtual screen with 60Hz refresh rate: Box breathing visualization (4-4-4-4 pattern) and 4-7-8 progressive relaxation technique. Both implementations utilize dynamic visual feedback.
**Creative Expression Module:** The Meadow environment features an interactive artistic interface. Users engage with virtual canvas, 3D stroke rendering with 6-color palette and two brush width.

This final implementation represents a significant advancement from previous prototypes, addressing both technological limitations and user-identified needs. By integrating high-fidelity environmental rendering, adaptive sensory calibration, and multi-modal interaction paradigms, the system offers an evidence-based approach to immersive relaxation experiences specifically optimized for neurodivergent users. The full Unity project is available on GitHub[6].

---
[6]https://github.com/kingaskiers/Portable-Silent-Room

| ID | Age | G | VR Exp. | Dx | Dx Method |
|---|---|---|---|---|---|
| P1 | 26–34 | F | YES | ADHD | Diagnosed |
| P2 | 26–34 | F | YES | ADHD | Diagnosed |
| P3 | 18–25 | NB | YES | Autism | Diagnosed |
| P4 | 35–39 | F | NO | ADHD | Diagnosed |
| P5 | 26–34 | F | NO | Autism | Diagnosed |
| P6 | 26–34 | F | YES | AuADHD | Diagnosed |
| P7 | 26–34 | F | NO | ADHD | Self-reported |
| P8 | 26–34 | F | YES | ADHD | Diagnosed |
| P9 | 18–25 | NB | YES | ADHD | Diagnosed |
| P10 | 35–39 | NB | NO | ADHD | Diagnosed |
| P11 | 50+ | NB | NO | AuADHD | Diagnosed |
| P12 | 50+ | NB | YES | AuADHD | Childhood Dx |
| P13 | 50+ | NB | NO | AuADHD | Diagnosed |
| P14 | 26–34 | F | NO | ADHD | Childhood Dx |
| P15 | 18–25 | F | YES | ADHD | Diagnosed |
| P16 | 26–34 | NB | NO | ADHD | Diagnosed |
| P17 | 18–25 | NB | YES | AuADHD | Diagnosed |
| P18 | 26–34 | F | YES | ADHD | Self-reported |
| P19 | 26–34 | F | NO | AuADHD | Diagnosed |
| P20 | 26–34 | F | YES | ADHD | Self-reported |
| P21 | 35–39 | NB | NO | ADHD | Diagnosed |
| P22 | 18–25 | F | YES | ADHD | Diagnosed |
| P23 | 26–34 | NB | NO | ADHD | Self-reported |
| P24 | 26–34 | F | NO | ADHD | Diagnosed |
| P25 | 26–34 | NB | YES | AuADHD | Diagnosed |

Table 2: Participant demographics. G = Gender (F: Female, NB: Nonbinary). Dx = Diagnosis. AuADHD=Autism & ADHD

### 6.2 Methods

#### 6.2.1 Participants

We recruited 25 participants (14 female, 11 non-binary) through autism and ADHD advocacy organizations, and online neurodivergent communities. This recruitment strategy addressed the inherent challenges of reaching ND populations, especially those marginalized by gender. Seventeen participants (68%) held formal clinical diagnoses of ADHD, Autism, or AuADHD (co-occurring autism and ADHD), while 8 participants (32%) did not. Among those without formal documentation, two were undergoing diagnostic evaluation during the study and later received clinical diagnoses, two had received childhood assessments. Four were self-diagnosed by The Autism Spectrum Quotient (AQ) and DSM-5 [74], citing financial or language-related barriers to accessing healthcare. Participants ranged in age from 18 to over 50. The majority (56%) were between 26–34 years old, 20% aged 18–25, 12% aged 35–39, and 12% aged 50 or older. Thirteen participants (52%) reported having prior experience with virtual reality systems, while twelve (48%) did not. This sample reflects diverse lived experiences across diagnostic status, age, and gender identity, representing the diversity of potential users of neurodivergent-centered technologies.

#### 6.2.2 Procedure

To validate the system's effectiveness, we implemented a distributed testing protocol that allowed participants to choose their own testing environments. This approach addressed potential stigma concerns around VR use in public spaces by letting participants select locations where they felt comfortable and secure. **Location Selection and Distribution**: Participants chose their preferred testing environments from four categories: university settings, café environments, home environments, and LGBTQ+ friendly social spaces. The distribution was 9 participants in university settings, 4 in cafés, 3 at home, and 9 in LGBTQ+ friendly spaces. This participant-driven selection ensured testing across diverse real-world contexts while minimizing discomfort or unwanted attention.
**Data Collection and Consent**: Upon arrival, participants completed informed consent acknowledging motion sickness risks and withdrawal rights, then answered a background survey covering demographics, neurodivergent status, current relaxation strategies, and anxiety management challenges. Participants were introduced to the VR system and allowed to freely explore the virtual environment without time restrictions. Post-experience data collection included a comprehensive survey and semi-structured interview evaluating overall experience, emotional regulation efficacy, comparative advantages, desired improvements, and likelihood of daily use.
**Addressing Public VR Use Concerns**: During pre-study interviews, participants were specifically asked about their comfort level using VR in their chosen spaces and whether they experienced concerns about drawing attention or feeling stigmatized. Notably, no participants reported embarrassment or discomfort related to using the VR headset in their selected environments.

#### 6.2.3 Data Analysis

The procedure for the data analysis followed the same highlighted in Section 5.2.3. Interviews were transcribed by the first author and together with written qualitative responses to written questionnaire analyzed using a inductive thematic analysis approach [8].

### 6.3 Results

Participants accounts documented the distinct challenges they faced in anxiety management, sensory processing, and emotional regulation. Their responses highlighted several intersectional factors: gender-specific social expectations conflicting with neurodivergent traits, heightened stigmatization when displaying non-normative behaviors, cumulative physiological and psychological effects of chronic masking and systemic barriers to appropriate accommodations in various contexts. Multiple participants (P3, P5, P7, P15, P16, P19, P22) (28%) specifically noted receiving inadequate support due to having their sensory needs mischaracterized as being "too emotional" or "overreactive" particularly in professional environments. Participants reported employing diverse self-regulation approaches prior to engaging with our system:
**Social Strategies:** Strategic disengagement and controlled social withdrawal
**Sensory Management:** Noise-canceling technologies and audio
**Cognitive Techniques:** Structured self-talk and imaginative immersion
**Physical Interventions:** Comfort object utilization and animal-assisted regulation
**Movement Regulation:** Rhythmic movement patterns and proprioceptive stimulation (e.g stimming)
**Creative Expression:** Various artistic modalities for emotional processing such as drawing, painting and writing
**Environmental Modifications:** Routine maintenance and space customization
Regarding specific regulation tools, participants frequently mentioned acoustic isolation devices, meditation practices, breathing techniques, and deep pressure stimulation products. Access to dedicated quiet spaces, varied substantially among participants: 24% reported regular access with frequent utilization, 32% had access but inconsistent usage, 40% lacked access despite identified need, and 4% indicated no requirement for such spaces. When it came to the evaluation of the portable Silent Room, participants widely appreciated the ability to personalize their virtual environment. The day/night toggle and virtual location selection were reported as the most useful features by 79.1% of respondents. Volume control was described as useful by 37.5% of participants, particularly when used in conjunction with calming background music. Several participants (P3, P8, P11, P13, P14, P22) (24%) emphasized breathing exercises and painting functionalities as exceptionally beneficial, describing them as "*the most relaxing*" and noting their parallel to real-life coping strategies. The customization of visual and auditory elements was frequently cited as enhancing immersion and reducing external stressors. One participant reflected: (P8)"*At first the night scene felt anxiety-inducing, but after I turned on white noise and started drawing, I felt safe enough to switch to night mode. Drawing is how I relax in real life.*"

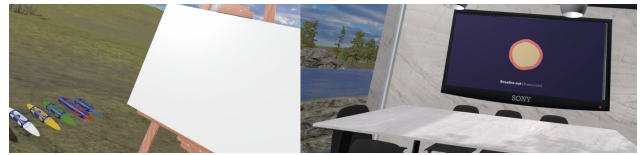

Fig. 5: Final prototype, Left: Painting, Right: Breathing Exercises

### 6.3.1 Desired Features and Improvements

Participants suggested several improvements that would enhance the system's capacity to support emotional regulation:

- **Sensory and comfort adjustments:** Participants requested expanded paint color options, including additional interactive elements (e.g., animals, instruments), and enhanced scene depth (e.g., visible seating areas, ambient candles). Some reported a sense of oppression in indoor environments and preferred scenes with open spaces. Three participants (12%) also mentioned a feeling of discomfort due to the headset's weight during extended sessions.
- **Interface and usability:** Five participants (20%) who were novice VR users noted challenges with hand tracking and pinch controls, suggesting enhancements such as toggleable hot bars, more precise drawing tools, and looped audio tracks to improve the user experience.
- **Mental health support:** Five participants (20%) recommended integrated mindfulness features such as introspection reminders, voice guidance options, or subtle prompts to support engagement during sessions.

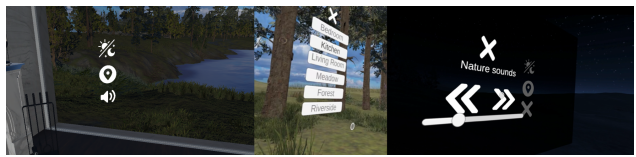

Fig. 6: Final prototype, From left: 3 UI toggles, Locations panel, audio panel

### 6.3.2 Potential for Daily Use

Participants expressed predominantly positive sentiment regarding regular system usage, with 83.3% indicating they would likely incorporate it into their daily routines if accessible. Primary barriers included current VR hardware cost and size constraints, with some participants noting they already maintain physical safe spaces. However, many highlighted VR's potential as a portable solution for environments where privacy or sensory regulation is challenging, such as workplaces or public settings:(P2)"*Sometimes there are situations where I can't just leave if I'm overwhelmed or overstimulated, but I need to regulate right away I would use VR for that.*"

Despite three individuals (12%) reporting discomfort from the headset's weight, six participants (24%) mentioned that the physical sensation of the headset (pressure on the head and ears) contributed positively to their calming experience, likening it to deep pressure therapy.

### 6.3.3 Relevance for Neurodivergent Users

Almost all participants (91.7%) affirmed that VR systems specifically designed for neurodivergent users are either necessary or highly beneficial. Key justifications included: The need for customized sensory environments, personalized emotional regulation tools, creating accessible safe spaces when physical alternatives are unavailable: (P24)"*We have more specific needs than neurotypical people when it comes to light, sounds, and sensory input. It's important to carefully design systems for us.*"

One participant advocated for universal accessibility integration in general VR applications rather than separate systems, alongside broader institutional support for sensory-safe physical spaces.

### 6.3.4 Suitable Environments for Deployment

Participants identified strong potential for system implementation across diverse real-world contexts. This includes workplaces (83.3%), public relaxation spaces (83.3%), therapy clinics (79.2%), educational institutions (54.2%) and home environments (54.2%). The importance of accessibility (62.5%) was consistently emphasized, particularly in high-stress, overstimulating environments such as offices, urban settings, and classrooms. Participants strongly endorsed the concept of VR regulation spaces in shared or public environments:

(P12)"*In the same way rest spaces are provided for physical issues, mental health spaces could be very beneficial. This could be that space a virtual one.*"

Several participants envisioned VR 'calm pods' in workplaces and schools, offering short immersive breaks to reduce overload and boost productivity.

### 6.3.5 Motion Sickness and Comfort

A noteworthy finding was the absence of motion sickness among most participants, including those who had previously avoided VR due to discomfort. Despite common concerns about vestibular sensitivity in VR environments, particularly among neurodivergent users, the system design proved unexpectedly comfortable. Participants frequently expressed surprise at their lack of discomfort, with comments such as (P16)"*I usually can't stay in VR for more than a few minutes, but this was fine*" and (P9)"*I was expecting to feel dizzy, but I didn't at all.*" This suggests that key design elements, including limited artificial locomotion, environmental stability, and gentle pacing, contributed significantly to minimizing sensory conflict and improving accessibility. The design intentionally avoided aggressive teleportation or rapid movement, instead employing a seated position to further reduce physical discomfort. The system also incorporated low-contrast transitions to prevent jarring visual experiences, while soft audio helped anchor users spatially, reducing disorientation. These findings suggest that calm, stable virtual environments prioritizing user comfort can provide viable alternatives to traditional VR experiences for users typically excluded due to motion-related side effects.

## 7 DISCUSSION

By providing a calming, customizable virtual experience, our system offers an accessible emotional regulation tool for sensory overload or heightened emotional states [65]. Results showed high user engagement, with 83.3% of participants willing to incorporate the system into daily routines. Participants particularly valued the immersive soundscapes that enhanced spatial awareness and provided emotional anchoring within the virtual space. These results suggest that our multi-sensory approach represent a key design consideration for creating emotionally safe VR environments for neurodivergent users. The day/night toggle and environment selection features (valued by 79.1% of participants) demonstrated the importance of environmental control for emotional regulation. The unexpected therapeutic benefit of the headset's physical pressure, likened to deep pressure therapy, suggests promising avenues for integrating tactile elements. Most significantly, no participants experienced motion sickness, including those who had previously abandoned VR due to discomfort [6, 15]. This can be attributed to several design decisions: stable environmental elements with minimal artificial movement, a predominantly seated interaction model, low-contrast visual transitions, and spatial audio anchoring. These results challenge prevailing assumptions about VR's limitations for neurodivergent users with heightened sensory sensitivity [5, 40, 42] and suggest that carefully crafted VR experiences can successfully overcome traditional barriers to accessibility. Participants identified multiple high-potential implementation contexts, with workplaces (83.3%), public relaxation spaces (83.3%), and therapy clinics (79.2%) emerging as primary candidates. This distribution reflects the system's versatility and addresses a critical need for on-demand emotional regulation tools in environments where traditional accommodations are limited. The concept of "VR calm space" in institutional settings received strong support, indicating a potential shift in how organizations might approach neurodivergent accommodations. As P12 noted, systems like the Portable Silent Room could become valuable tools for sensorial and emotional rest which can easily be made accessible in a variety of public spaces.

In centering the experiences of women and non-binary neurodivergent individuals, groups historically underrepresented in both clinical and HCI research, this study brings important new insights into how gendered experiences shape emotional regulation needs. Participants frequently described masking, social stigma, and culturally embedded expectations as contributing factors to heightened anxiety and sensory overload [18, 58]. One non-binary participant (P17) noted, *"The level of scrutiny women and AFAB (assigned female at birth) people feel in the event they dare display any traits seen as 'unwomanly' is ineffable...that is doubly experienced by neurodivergent, and especially autistic women and AFABs."*, in the other hand, a female participant (P14) reflected,*"There is a social stigma that women are "too emotional" by nature, so they do not get proper help, it is all blamed on the women's nature"*. It was further emphasized about intersection of gender and sensory overwhelm by different participant (P25): *"As a non-binary person, I'm suffering a lot with sensory overload... the world expects us to mask constantly, and it's exhausting"*. These responses highlight how emotional dysregulation is not merely a sensory issue but often deeply entangled with gendered social pressures. [26, 58]. Many existing VR systems emphasize task-driven performance, yet research shows women report lower engagement in such environments compared to men [17]. Our system removes mandatory tasks and instead offers optional, low-pressure activities like breathing exercises, painting, or passive music listening. Traditional systems favor high-stimulation, goal-oriented interactions that align with cis-male preferences [38] but may overwhelm users with sensory sensitivities. Women experience higher rates of VR motion sickness [45, 71], while research on non-binary users remains virtually unexplored, we developed a system centered on user choice, sensory control, and gentle interaction through co-design with women and non-binary participants to address this gap in inclusive VR design.

Based on participant feedback, we propose several design considerations for future neurodivergent-centered VR systems: enhanced customization for sensory control, integrated mindfulness tools with optional guided practices, improved interaction design to reduce cognitive load, and adaptive experiences responsive to users' emotional states. While our findings demonstrate strong potential, several limitations warrant attention. Our sample of 25 participants, though neurodivergently diverse, limits generalizability, and our evaluation focused on immediate experience rather than long-term outcomes. Future research should investigate longitudinal effects, accessibility for users with lower digital literacy or physical differences, and integration of biometric feedback for real-time adaptation to user anxiety levels. Large-scale studies involving both neurodivergent and neurotypical users would strengthen external validity and inform more inclusive VR design. Overall, our findings lay the groundwork for adaptive, personalized immersive systems and emphasize the importance of centering neurodivergent perspectives throughout the design process.

## 8 CONCLUSION

The Portable Silent Room demonstrates potential as an accessible, effective tool for supporting emotional regulation among neurodivergent individuals. The system's strong acceptance among participants challenges conventional limitations of VR accessibility for neurodivergent users, particularly regarding motion sickness and sensory comfort. By prioritizing user-centered design principles and sensory accommodation, our research contributes valuable insights to the growing field of therapeutic XR applications. Participants positive attitudes towards potential deployment across various institutional contexts suggests that such systems could help address the critical shortage of accessible regulation spaces for neurodivergent individuals. As VR technology becomes increasingly portable and affordable, systems like ours may offer unprecedented opportunities for on-demand emotional support in environments previously inaccessible to neurodivergent individuals requiring sensory accommodation.

## ACKNOWLEDGMENTS

This work is supported by JST COI-NEXT "Minds1020Lab" Project Grant Number JPMJPF2203. This work is also partially funded by the University of Auckland Faculty of Science Research Development Fund Grant Number 3731533.